\documentclass[journal]{IEEEtran}
 \usepackage{amsmath,amssymb}
 \usepackage{subfigure}
 \usepackage{graphicx,graphics,color,psfrag}
 \usepackage{cite,balance}
 \usepackage{caption}
 \allowdisplaybreaks
 \usepackage{algorithm}
 \usepackage{accents}
 \usepackage{amsthm}
 \usepackage{bm}
 \usepackage{algorithmic}
 \usepackage[english]{babel}
 \usepackage{multirow}
 \usepackage{enumerate}
 \usepackage{cases}
 \usepackage{stfloats}
 \usepackage{dsfont}
 \usepackage{color,soul}
 \usepackage{amsfonts}
 \usepackage{cite,graphicx,amsmath,amssymb}
 \usepackage{subfigure}
 \usepackage{fancyhdr}
 \usepackage{hhline}
 \usepackage{graphicx,graphics}
 \usepackage{array,color}
 \usepackage{amsmath}
 \usepackage{booktabs}

\def\phi{\varphi}

\def\({\left(}
\def\){\right)}

\def\b0{{\mathbf{0}}}

\begin{document}
\setlength{\topskip}{-3pt}

\title{\huge Integrating Sensing, Communication, and Power Transfer: From Theory to Practice}
\author{Xiaoyang Li, Zidong Han, Guangxu Zhu, Yuanming Shi, Jie Xu, Yi Gong, Qinyu Zhang, Kaibin Huang, and Khaled B. Letaief
\thanks{Xiaoyang Li and Guangxu Zhu are with the Shenzhen Research Institute of Big Data, The Chinese University of Hong Kong-Shenzhen, Guangdong, China. Zidong Han and Qinyu Zhang are with Harbin Institute of Technology, Shenzhen, China. Yuanming Shi is with ShanghaiTech University, Shanghai, China. Jie Xu is with The Chinese University of Hong Kong, Shenzhen, China. Yi Gong is with Southern University of Science and Technology, Shenzhen, China. Kaibin Huang is with The University of Hong Kong, Hong Kong. Khaled B. Letaief is with The Hong Kong University of Science and Technology, Hong Kong.} 
}
\maketitle


\begin{abstract}
To support the development of internet-of-things applications, an enormous population of low-power devices are expected to be incorporated in wireless networks performing sensing and communication tasks. As a key technology for improving the data collection efficiency, \emph{integrated sensing and communication} (ISAC) enables simultaneous data transmission and radar sensing by reusing the same radio signals. In addition to information carriers, wireless signals can also serve as energy delivers, which enables \emph{simultaneous wireless information and power transfer} (SWIPT). To improve the energy and spectrum efficiency, the advantages of ISAC and SWIPT are expected to be exploited, leading to the emerging technology of \emph{integrating sensing, communication, and power transfer} (ISCPT). In this article, a timely overview of ISCPT is provided with the description of the fundamentals, the characterization of the theoretical boundary, the discussion on the key technologies, and the demonstration of the implementation platform.
\end{abstract}

\section{Introduction}
The \emph{sixth generation} (6G) of wireless networks are anticipated to offer a variety of \emph{internet-of-things} (IoT) applications, such as smart logistics, smart city, industrial automation, and smart home \cite{shi2023task}. To support these applications, billions of low-power IoT devices are expected to be incorporated in wireless networks performing sensing and communication tasks. Different from the traditional systems where the sensing and communication processes are independently constructed in different systems over different frequency bands, the \emph{integrated sensing and communication} (ISAC) system transmits unified signals for both data transmission and radar sensing in the same frequency band to increase the spectrum utilization efficiency and facilitate the data collection \cite{liu2021integrated}. 

In addition to information carriers, wireless signals can also serve as energy delivers. As indicated by the Koomey's law, the number of low-power devices will exponentially increase with the reducing power requirements of the electronic devices \cite{koomey2010implications}. \emph{Wireless power transfer} (WPT), which uses microwave signals to transmit energy, is anticipated to supply the energy for these low-power devices \cite{zeng2017communications}. In a variety of contexts, such as mobile edge computing, fast data aggregation, mobile crowd sensing, and ISAC, WPT has been widely used to power the devices as a replacement of the charging cables. The practical experiments have verified that WPT can charge a variety of low-power wearables and mobile devices.

\newpage
By integrating the deliveries of information and energy, \emph{simultaneous wireless information and power transfer} (SWIPT) has been proposed \cite{clerckx2018fundamentals}. SWIPT leverages the principles of microwave waves to transfer power and information via the same signals at the same time. This eliminates the need for separate power sources or wired connections, leading to improved flexibility, mobility, and convenience. SWIPT has the potential to revolutionize various industries and applications, including wireless sensor networks, IoT, wearable devices, and autonomous systems. However, most prior works on SWIPT concentrate on using wireless signals as power and communication conduits, seldom leveraging the sensing capabilities of the signals.

To account for the explosive growth of the number of IoT devices and the resultant dense network, it is necessary to further improve the energy and spectrum efficiency. An initial study came up with the idea of \emph{integrating sensing and wireless power transfer} (ISWPT) \cite{yang2022beamforming}. As a further step, the \emph{integrating sensing, communication, and power transfer} (ISCPT) technology enables simultaneous sensing along with information and energy delivery in the same frequency band via proper signal design. The harvested energy via ISCPT can be used for sensing and communication activities of low-power devices. Compared with the conventional approaches, ISCPT can improve the utilization efficiency of radio resources. Therefore, ISCPT is expected to support the sustainable operation of a series of applications such as low-power radars and \emph{radio frequency} (RF) identification devices.

In this article, an overview of ISCPT is provided with the introduction of the fundamentals, the characterization of the theoretical boundary, the discussion on the key techniques, and the demonstration of the implementation platform. To the best of the authors’ knowledge, this is the first overview article on ISCPT.

\begin{figure*}[ht]
\centering
\includegraphics[scale=0.45]{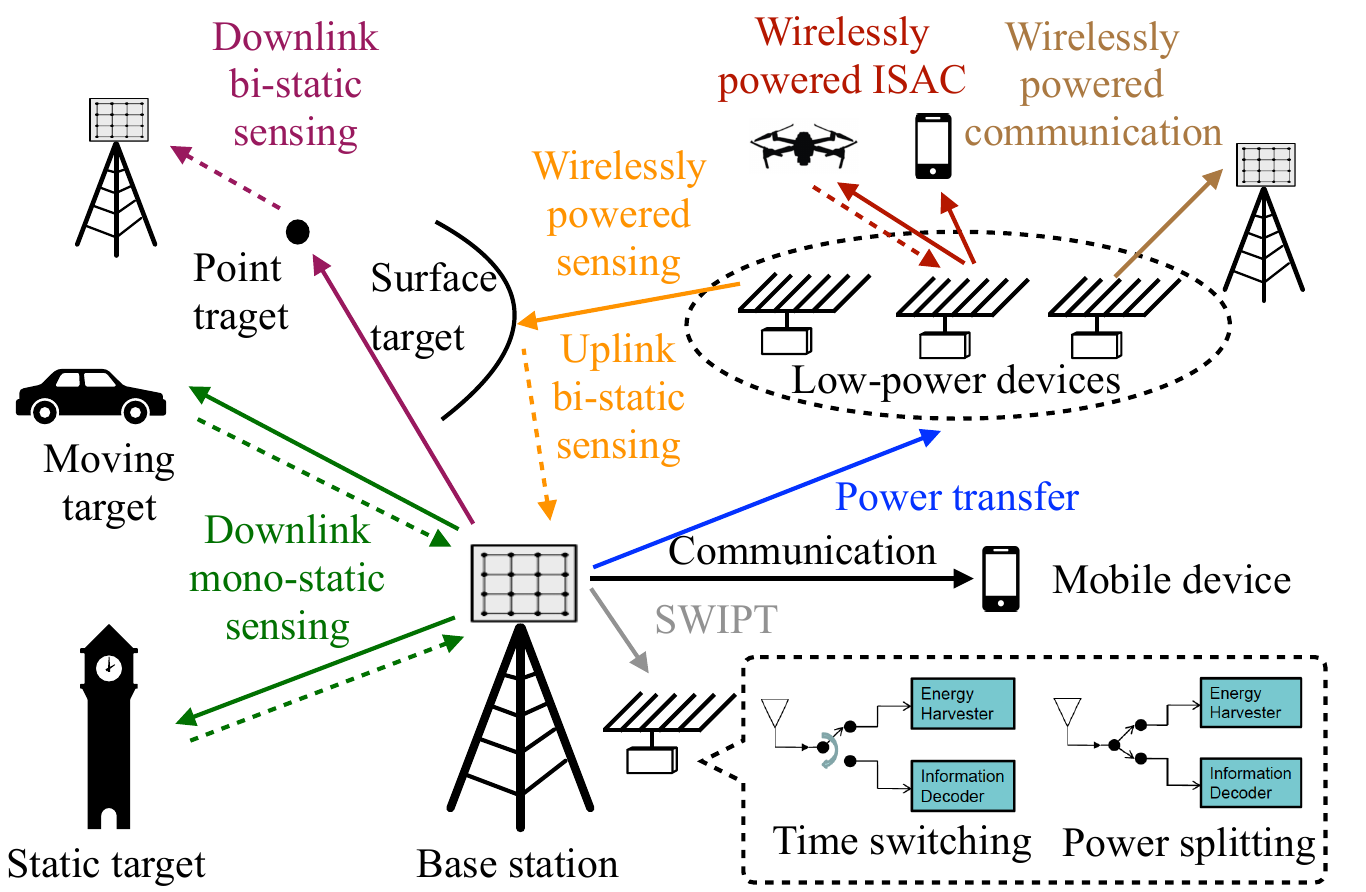}
\caption{ISCPT framework.}
\label{FigFramework}
\end{figure*}

\section{ISCPT Fundamentals}
\subsection{ISAC}
ISAC is a cutting-edge technology that combines the functionalities of sensing and communication within the same system. Unlike traditional approaches where sensing and communication systems operate independently, ISAC utilizes a shared frequency band to generate unified signals for both data transmission and radar sensing. This integration offers numerous benefits, including improved spectrum efficiency, simplified data collection processes, and the ability to support a wide range of sensing and communication services. By seamlessly merging these capabilities, ISAC paves the way for advanced applications in the next-generation networks, and enables efficient and versatile operations in various frequency bands, ranging from the S-band to the terahertz band. 

Because of its benefits, ISAC has a wide range of applications. In vehicular networks, a pioneering work applies ISAC for real-time traffic monitoring, vehicle detection, and tracking \cite{yuan2021integrated}. In smart cities, ISAC enables real-time monitoring of various aspects such as traffic, waste management, and energy consumption. As for industrial automation, ISAC integrates sensors, actuators, and communication systems to enable seamless data exchange and coordination. As for healthcare, ISAC enables the integration of medical sensors and communication devices to facilitate real-time monitoring of patients and improve the healthcare efficiency. ISAC can also assist the security and surveillance systems by enabling real-time monitoring of sensitive areas, object detection, tracking, and communication between surveillance devices. Despite the widely applications of ISAC, the utilization of energy carried by the ISAC signals remains unexplored.

\subsection{WPT}
WPT enables the transmission of electrical power from a power source to a recipient device without the need of physical wired connections. It offers the convenience of wire-free power delivery, eliminating the need for cables to charge the devices. The methods for supporting WPT systems include inductive coupling, resonant coupling, and RF transmission \cite{kurs2007wireless}. Inductive coupling involves the use of coils or transformers to create a magnetic field that induces an electric current in the recipient device. Resonant coupling utilizes resonant circuits to transfer power efficiently between a power source and a receiver in the same resonant frequency. RF-based WPT uses radio waves to transmit power through the air, which is then captured and converted into electrical power by the energy harvesting devices. While WPT technology offers convenience and flexibility, it also faces challenges such as efficiency, distance limitations, alignment requirements, safety considerations, and compatibility with existing devices and standards. The ongoing research and development efforts are focused on improving these aspects to enhance the efficiency and reliability of WPT systems.

\subsection{SWIPT}
As a cutting-edge technology, SWIPT enables the concurrent transmission of both data and electrical power wirelessly. The information and power receiving can be conducted in time-switching or power-splitting manners \cite{zhang2013mimo}. In time-switching manner, the signal receiver alternates between the energy harvesting and information decoding modes during different time intervals. The time allocation between power transfer and information transmission can be predetermined or dynamically adjusted based on system conditions and requirements. The duration of each time slot can vary depending on factors such as the energy harvesting capabilities, the power requirements of the device, and the data communication demands. Power splitting is another essential technique used in SWIPT systems. It divides the received signal into two parts for energy harvesting and information decoding, respectively. This technique enables the simultaneous energy harvesting and information decoding. 

\subsection{ISWPT}
It should be noted that the energy is carried not only by communication signals but also sensing signals. Therefore, the functionalities of sensing and power transfer can also be naturally combined. In ISWPT, the radar sensing and wireless power transfer functionalities are integrated into one hardware platform \cite{yang2022beamforming}. The integration brings several benefits with respect to spectrum efficiency, energy consumption, and hardware costs. There is a tradeoff between the sensing performance and power transfer performance, which is affected by the transmitted signal and beamforming design. According to the practical requirements, these parameters can be adjusted respectively. 

\subsection{ISCPT}
By combining the functionalities of sensing, communication, and power transfer, we have the ISCPT framework as depicted by Fig. 1. The implementation of ISCPT has different modes. First, the wireless signals can be used for triple functions of powering devices, sensing targets, and passing messages to users at the same time. Next, ISCPT also enables a new application of wireless powered ISAC, where the \emph{base station} (BS) can use WPT to charge low-power devices, which then use the harvested energy for performing individual ISAC tasks. The ISCPT can be implemented in both narrow band and wide band \cite{chen2022isac}. The ISCPT functionalities can be realized by \emph{Orthogonal Frequency Division Multiplexing} (OFDM) signals \cite{zhang2023multi}. The power transfer happens in near field, while the sensing and communication functionalities can be achieved in far field. The sensing target can be regarded as a surface, a point, a static object, or a moving object.

In ISCPT, the power transfer is usually a downlink process from the BS to wireless devices, while the information transfer process can be either on the downlink or uplink. The sensing functionalities can be downlink mono-static, downlink bi-static, and uplink bi-static. Since the sensing, communication and power transfer have different objectives, the tradeoff between these functionalities becomes essential. How to optimally balance such tradeoffs under different configurations becomes interesting problems worth pursuing.

\section{Theoretical Boundary of ISCPT}
To illustrate the tradeoff among the performances of sensing, communication, and power transfer, our initial study focused on charactering the theoretical boundary of ISCPT in the multi-functional MIMO system \cite{chen2022isac}. In this framework, a multi-antenna BS transmits signals to simultaneously power an \emph{energy receiver} (ER), deliver message to an \emph{information receiver} (IR), and sense a target. The tradeoff among the performances of sensing, communication, and power transfer for such a multi-functional wireless system depends on the transmit design at the BS, which leads to a challenging optimization problem. 

First, the objectives of sensing, communication, and power transfer are in distinct forms, thus leading to distinct transmit design principles. As for MIMO radar sensing, the isotropic transmission based on an identity sample covariance matrix can exploit the waveform diversity to improve the sensing performance. As for MIMO communication, the eigenmode transmission with the water-filling power allocation over decomposed parallel sub-channels is known to be optimal to maximize the MIMO channel capacity. As for MIMO WPT, the strongest eigenmode transmission based on the WPT channel is generally preferred to maximize the received energy. Second, the inter-signal-stream interference has different effects on the sensing, communication, and power transfer performances. Specifically, the inter-signal-stream interference can be exploited to enlarge the harvested energy, but can be harmful for sensing and communication functionalities and thus needs to be mitigated. Due to these issues, it is difficult to find a unified transmit design to maximize the three objectives at the same time. How to balance the performance tradeoffs becomes thus important.

\begin{figure}[t]
\centering
\includegraphics[scale=0.6]{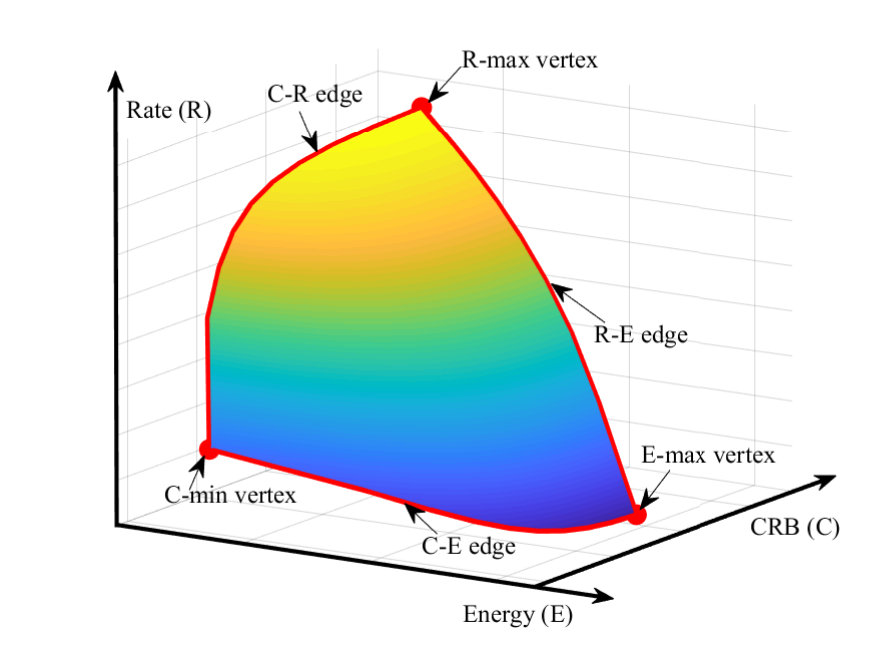}
\caption{Theoretical boundary of ISCPT \cite{chen2022isac}.}
\label{FigBoundary}
\end{figure}

To characterize such tradeoffs, we define the achievable ISCPT performance region for sensing, communication, and power transfer, and accordingly employ the corresponding Pareto boundaries to show the performance limits. In particular, we consider two different target models for radar sensing. If the target is regarded as a point, the sensing performance is evaluated by the \emph{Cram\'er-Rao bound} (CRB) of the target angle estimation error. If the target is regarded as an extended surface, the sensing performance is evaluated by the CRB for estimating the \emph{target response matrix} (TRM). The communication performance is evaluated by the achievable rate, and the power transfer performance is evaluated by the amount of harvested energy. As shown in Fig. 2, the three vertices of the Pareto boundary with respect to the sensing, communication, and power transfer functionalities are derived based on the minimization of the sensing CRB, the maximization of the achievable rate, and the maximization of the harvested energy. The remaining points on the Pareto boundary are characterized by solving the rate maximization problem under the constraints of tolerable sensing CRB, required amount of harvested energy, and the transmit power budget at the BS.

Fig. 2 illustrates the Pareto boundary of sensing, communication, and power transfer obtained in the initial study. It can be observed that the Pareto boundary corresponds to a curved surface in a three-dimensional space, where improving one performance metric will inevitably result in the deterioration of others. This illustrates the nonlinear tradeoff among the performances of sensing, communication, and power transfer. The Pareto boundary is affected by the number of receive antennas at the ER and IR, the transmit power, and the number of transmit antennas at the BS.

\section{Key ISCPT Technologies}
Recent researches on ISCPT focus on key technologies for powering wireless devices, enhancing radar sensing accuracy, and increasing communication throughput. To this end, a series of technologies have been proposed in our initial studies, including spatial multiplexing, multi-targets detection, and resource allocation.

\subsection{Spatial Multiplexing}
To power multiple devices, spatial multiplexing is required to support MIMO ISCPT. In our considered framework, the BS equipped with multiple antennas simultaneously powers multiple ERs, delivers message to multiple IRs, and senses a target. The functionalities of sensing, communication and power transfer are realized in one signal. The tradeoff among the performances of sensing, communication, and energy harvesting is reflected in the beamforming design. The sensing performance is evaluated by the CRB of the parameter estimation error. The communication performance is evaluated by the \emph{signal-to-interference-plus noise-ratio} (SINR) of the received signal at each IR. The power transfer performance is evaluated by the harvested energy at each ER. The locations of the ERs and IRs can be the same or different, which are known as the co-located case and separated case. In the separated case, ERs and IRs have different channel conditions. In the co-located case, the received signal is split for information receiving and energy harvesting. The beamforming designs together with the performance analysis are discussed in \cite{li2023multi}.

\begin{figure}[t]
\centering
\includegraphics[scale=0.45]{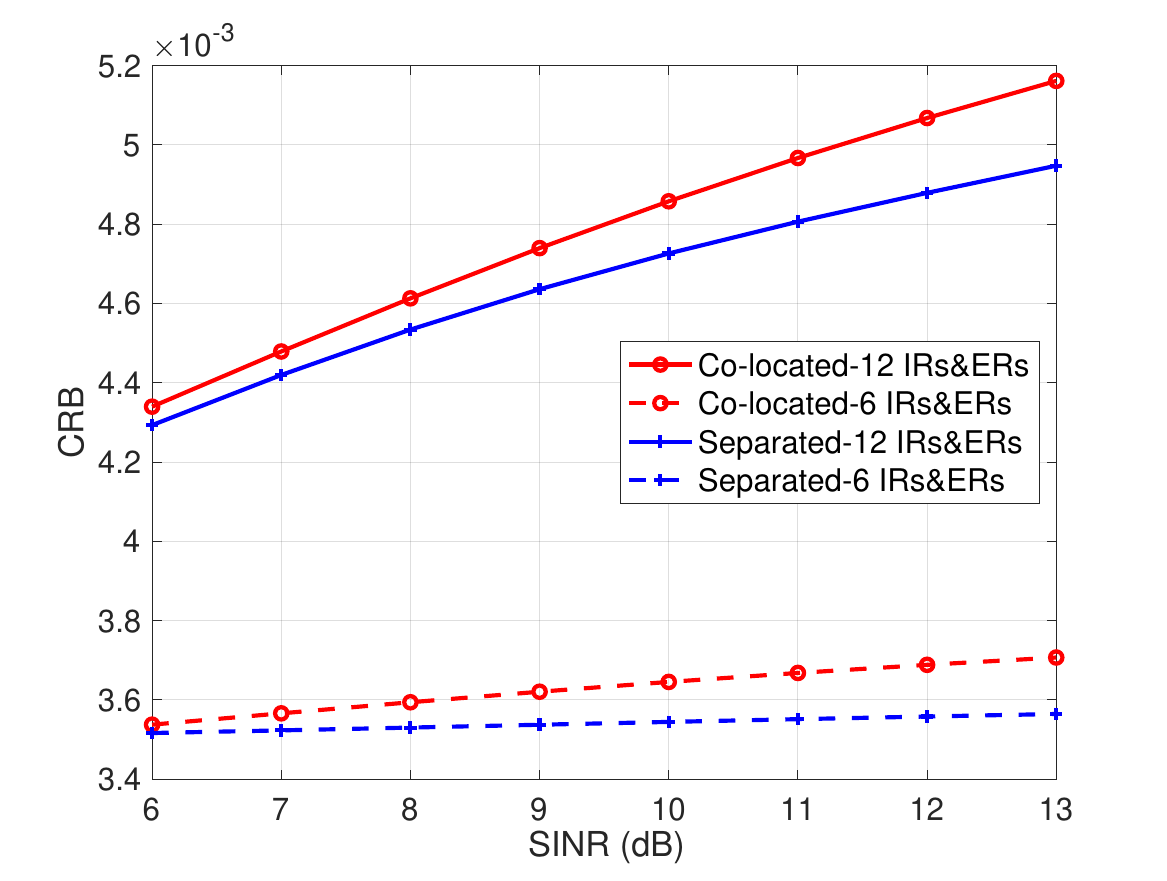}
\caption{Spatial multiplexing for ISCPT \cite{li2023multi}.}
\label{FigSpatial}
\end{figure}

\begin{figure}[t]
\centering
\includegraphics[scale=0.6]{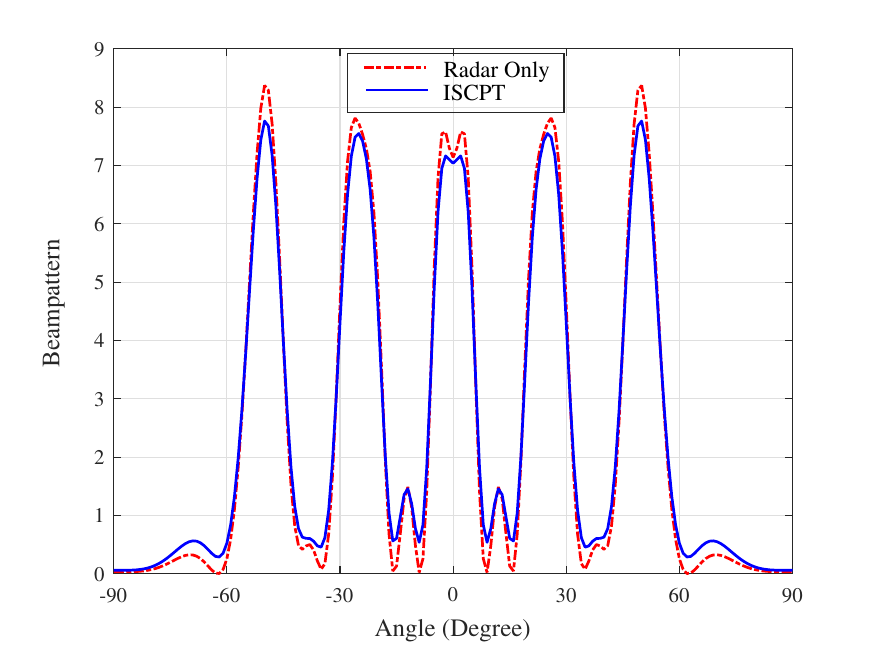}
\caption{Multi-targets detection via ISCPT \cite{zeng2022beamforming}.}
\label{FigMulti}
\end{figure}

Fig. 3 illustrates the performance comparison between the separated and co-located IR/ER cases under different settings of IR amounts. It can be observed that the sensing CRBs in both cases increase with the increasing of required SINR by IRs. This illustrates the tradeoff between the sensing and communication performances in ISCPT design. One can also observe that the sensing CRB in the co-located case is larger than that in the separated case. This is due to the effect of power splitting, which amplifies the influence of the SINR requirement. Moreover, when there are 12 IRs, the sensing CRB is larger compared with the case with 6 IRs. This illustrates that guaranteeing the SINR requirements of more IRs leads to the deteriorated sensing performance.

\subsection{Multi-targets Detection}
In practice, there might be multiple targets to be detected via ISCPT. A framework is considered in our initial study \cite{zeng2022beamforming}, where the BS equipped with a uniform linear array simultaneously senses multiple targets, passes messages to multiple IRs, and powers multiple ERs. The functionalities of communication and power transfer are realized via separated signals. Without loss of generality, each IR/ER is assigned a dedicated information/energy beam. All the antennas are shared for not only radar detection but also information and power transfer, which implies that the signals carrying information and energy are also used for sensing. Without considering the existence of IRs and ERs, the sensing-only beampattern design of MIMO radar is depicted in \cite{stoica2007probing}. Taking the information and power transfer functionalities into consideration will result in the deviation from the sensing-only beampattern. Such a deviation is known as the beampattern matching error, which can be minimized by proper beamforming design. The resultant non-convex problem can be solved by the difference-of-convex method.

Fig. 4 shows the performance comparison between the sensing-only beampattern and the beampattern for ISCPT. There are 5 peaks in the beampattern corresponding to the 5 sensing targets. It can be observed that though there exists a gap between the sensing-only beampattern and the beampattern for ISCPT, the beams for ISCPT can also point to the direction of the sensing targets. This shows that the proper beamforming design for ISCPT can effectively support multi-targets detection. However, the sensing accuracies of the specific parameters need to be further characterized.

\subsection{Resource Allocation}
To improve the performance of ISCPT, the allocation of resources such as power, spectrum, and time warrants further investigation. The power allocation for ISCPT is investigated in our initial study \cite{li2022wirelessly}. As depicted in Fig 5, a power beacon is used to charge multiple low-power devices. Based on the harvested energy, each device transmits ISAC signals for passing message and detecting targets. The sensing performance is evaluated by the \emph{mean squared error} (MSE) of the TRM and the communication performance is evaluated by the SINR. To improve the ISAC performance of each low-power devices, the power control, ISAC beamforming, and WPT beamforming are jointly optimized.

\begin{figure}[t]
\centering
\includegraphics[scale=0.45]{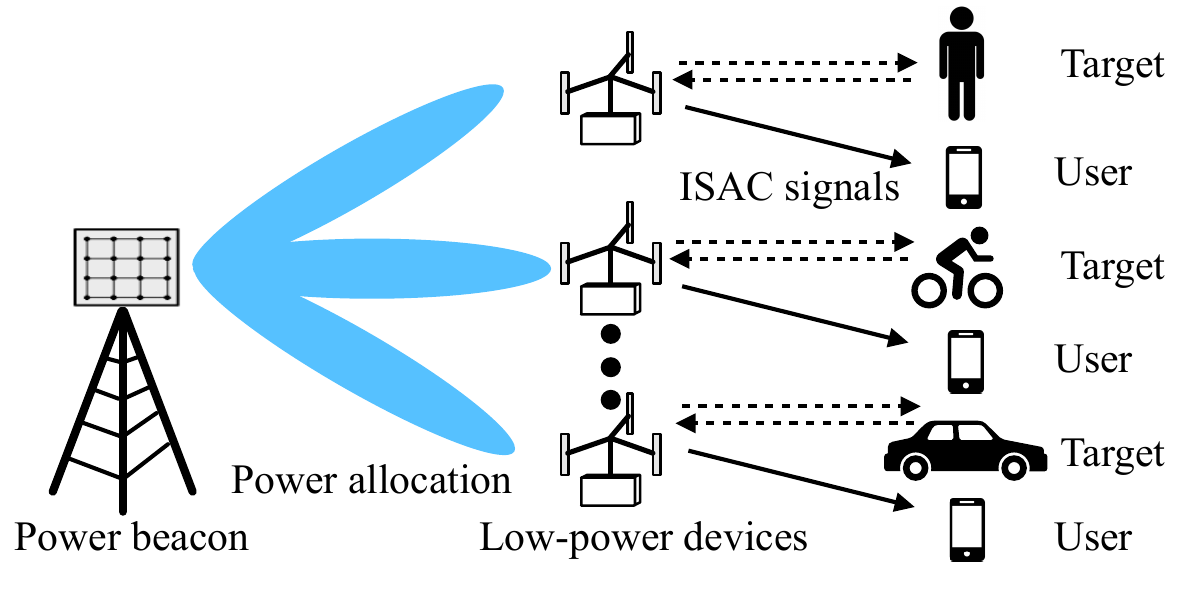}
\caption{Resource allocation for ISCPT \cite{li2022wirelessly}.}
\label{FigResource}
\end{figure}

Our study finds that increasing the power budget can effectively reduce the sensing error. Moreover, more transmit antennas at each device results in larger sensing error, as the dimension of TRM to be estimated is enlarged. It should be noted that this is an initial study on the power allocation for ISCPT, while the effects of other resource allocation on the ISCPT performance remains uncharted.

\section{Experimental Platform}
The ISCPT experimental platform is constructed to perform and evaluate wireless sensing, communication and power transfer via electronic measurement instruments. As shown in Fig. 6, the platform is composed of a Keysight Vector Signal Generator (marked by ‘A’), an 8-channel Beamformer (B), a Tektronix \emph{Direct Current} (DC) Power Supply (C), a 4x8 Transmit Antenna Array (D), a 1x4 Recieve Antenna Array (E), a RF-to-DC Energy Harvester (F), a Keysight High-Precision DC Power Analyzer (G), a Spectrum Analyzer (H), a RF Power Meter (I), a box stack as sensing target (J), a Horn antenna (K), and a National Instruments 2-channel Software Defined Radio (L).

In this platform, the working frequency is 2.45 GHz with 100 MHz bandwidth, i.e., 2.4 - 2.5 GHz. The ISCPT BS consists of A, B, C, D, K and L. The signals are produced by A, where the standard \emph{quadrature phase shift keying} (QPSK) and \emph{frequency modulated continuous wave} (FMCW) are chosen to perform communication and sensing respectively. They can be combined in various modes, such as time-divided, frequency-divided, multi-tone, spatial-divided, or more advanced signal waveform patterns that deserve deeper researches. The signals also carry energy for WPT. B is a beamformer with 8-channel RF power amplifier (output up to 2-watt in total) that can split the input signal into 8 steams with pre-designed individual amplitudes and phases. C is an antenna array for transmitting signal beam, with horizontal direction adjusted by B and vertical direction adjusted by tripod head. D works as the power supply of B. K works as the receive antenna of reflected sensing signals. L is connected with A, K, and an upper computer, used for simultaneously collecting the two streams of sensing signals (split by a Keysight two-way power splitter at A), i.e., one way is reflected by the targeting box stack, another is from A as the reference signal.

To evaluate the sensing functionality, L captures the FMCW signals reflected by J as well as the signals from A simultaneously to perform radar ranging. Before capturing the target signals, L should be calibrated to precisely calculate the time difference between the two RF channels. The Fourier transform of the \emph{intermediate frequency} (IF) signal is demonstrated in Fig. 6. The ranging results can be calculated based on the principle of radar ranging, e.g., the result of the Fourier transform of the IF signal. The accuracy could be influenced by various factors, such as the SINR and the signal processing algorithms. 

The evaluation framework of WPT is realized by F and G. F is a RF-to-DC energy harvester connected with a 4.2 V 100 mF super capacitor as an energy storage. The efficiency of RF-to-DC is usually fluctuating, influenced by the input signal power, radio propagation, channel condition and waveform patterns. Therefore, to stimulate chipsets as well as circuits, an energy storage after a DC-to-DC booster is still necessary until the power consumption of ultra-low power IoT chipsets reduces to an acceptable level in the near future with the rapid development of semiconductor science and engineering. The DC generated by the super capacitor is measured by G, a DC power analyzer equipped with a high-precision internal-resistance-compensated power measurement unit, which can detect nano-amps current. The voltage, current and power pulses of the received signal by the energy harvester are shown in Fig. 6. In this experiment setup, the transmit power is adjusted to be totally about 2 watts at the output of the beamformer. The power of the beam measured by I at E (the distance from D to E is about 2 meters) is averagely about 4 dBm with some random fluctuations. The DC pulse peak is about 4.17V/20.9mA with a 198.7 Ohm load resistance.

To evaluate the communication functionality, H, a 10 Hz - 7 GHz Spectrum Analyzer, works as a communication receiver which is equipped with a Keysight Signal Analysis Software, to perform and evaluate the digital demodulation. The real-time demodulation constellation of the received QPSK signal is illustrated in Fig. 6. Multiple communication performance metrics can be calculated in H, including error vector magnitude, SINR, phase offset and so on. 

\begin{figure*}[ht]
\centering
\includegraphics[scale=0.5]{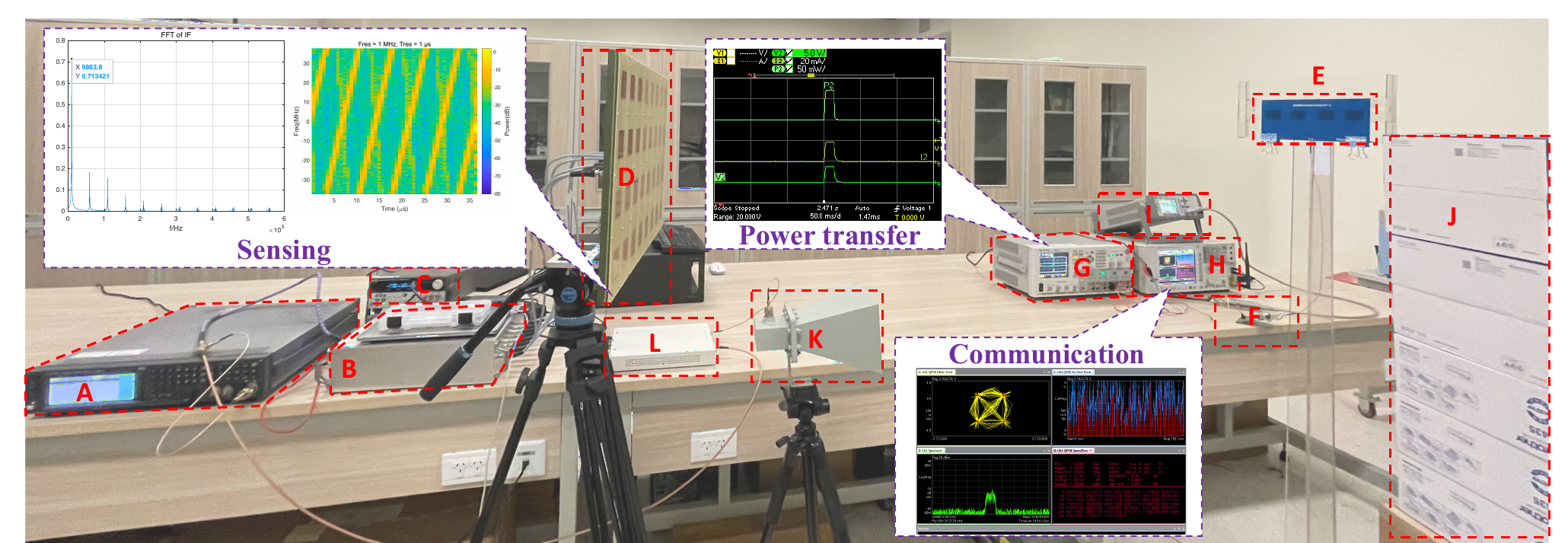}
\caption{ISCPT platform and functionalities.}
\label{FigPlatform}
\end{figure*}

\section{Concluding Remarks}
In 6G wireless networks, ISCPT is expected to enable simultaneous sensing, communication, and power transfer for the expected massive number of IoT devices. The fundamentals of ISCPT are introduced in this paper, followed by the theoretical boundary. To improve the performance of ISCPT, multiple key technologies are proposed, including spatial multiplexing, multi-targets estimation, and resource allocation. An implementation platform is further demonstrated to show the potential application of ISCPT in practice. This paper is intended to spur interests and further investigations on the future evolution of ISCPT.

\section*{Acknowledgment}
Thanks Prof. Zhisheng Niu from Tsinghua University for his advice on improving the quality of this work. 

\bibliographystyle{IEEEtran}
\bibliography{cite}

\begin{thebibliography}{10}
\providecommand{\url}[1]{#1}
\csname url@samestyle\endcsname
\providecommand{\newblock}{\relax}
\providecommand{\bibinfo}[2]{#2}
\providecommand{\BIBentrySTDinterwordspacing}{\spaceskip=0pt\relax}
\providecommand{\BIBentryALTinterwordstretchfactor}{4}
\providecommand{\BIBentryALTinterwordspacing}{\spaceskip=\fontdimen2\font plus
\BIBentryALTinterwordstretchfactor\fontdimen3\font minus
  \fontdimen4\font\relax}
\providecommand{\BIBforeignlanguage}[2]{{%
\expandafter\ifx\csname l@#1\endcsname\relax
\typeout{** WARNING: IEEEtran.bst: No hyphenation pattern has been}%
\typeout{** loaded for the language `#1'. Using the pattern for}%
\typeout{** the default language instead.}%
\else
\language=\csname l@#1\endcsname
\fi
#2}}
\providecommand{\BIBdecl}{\relax}
\BIBdecl

\bibitem{shi2023task}
Y.~Shi, Y.~Zhou, D.~Wen, Y.~Wu, C.~Jiang, and K.~B. Letaief, ``Task-oriented
  communications for {6G}: Vision, principles, and technologies,'' \emph{IEEE
  Wireless Commun.}, vol.~30, no.~3, pp. 78--85, 2023.

\bibitem{liu2021integrated}
F.~Liu, Y.~Cui, C.~Masouros, J.~Xu, T.~X. Han, Y.~C. Eldar, and S.~Buzzi,
  ``Integrated sensing and communications: {Towards} dual-functional wireless
  networks for {6G} and beyond,'' \emph{IEEE J. Sel. Areas Commun.}, vol.~40,
  no.~6, pp. 1728--1767, 2022.

\bibitem{koomey2010implications}
J.~Koomey, S.~Berard, M.~Sanchez, and H.~Wong, ``Implications of historical
  trends in the electrical efficiency of computing,'' \emph{IEEE Ann. Hist.
  Comput.}, vol.~33, no.~3, pp. 46--54, 2010.

\bibitem{zeng2017communications}
Y.~Zeng, B.~Clerckx, and R.~Zhang, ``Communications and signals design for
  wireless power transmission,'' \emph{IEEE Trans. Commun.}, vol.~65, no.~5,
  pp. 2264--2290, 2017.

\bibitem{clerckx2018fundamentals}
B.~Clerckx, R.~Zhang, R.~Schober, D.~W.~K. Ng, D.~I. Kim, and H.~V. Poor,
  ``Fundamentals of wireless information and power transfer: {From RF} energy
  harvester models to signal and system designs,'' \emph{IEEE J. Sel. Areas
  Commun.}, vol.~37, no.~1, pp. 4--33, 2018.

\bibitem{yang2022beamforming}
Q.~Yang, H.~Zhang, and B.~Wang, ``Beamforming design for integrated sensing and
  wireless power transfer systems,'' \emph{IEEE Commun. Lett.}, vol.~27, no.~2,
  pp. 600--604, 2022.

\bibitem{yuan2021integrated}
W.~Yuan, Z.~Wei, S.~Li, J.~Yuan, and D.~W.~K. Ng, ``Integrated sensing and
  communication-assisted orthogonal time frequency space transmission for
  vehicular networks,'' \emph{IEEE J. Sel. Topics Signal Process.}, vol.~15,
  no.~6, pp. 1515--1528, 2021.

\bibitem{kurs2007wireless}
A.~Kurs, A.~Karalis, R.~Moffatt, J.~D. Joannopoulos, P.~Fisher, and
  M.~Soljacic, ``Wireless power transfer via strongly coupled magnetic
  resonances,'' \emph{science}, vol. 317, no. 5834, pp. 83--86, 2007.

\bibitem{zhang2013mimo}
R.~Zhang and C.~K. Ho, ``{MIMO} broadcasting for simultaneous wireless
  information and power transfer,'' \emph{IEEE Trans. Wireless Commun.},
  vol.~12, no.~5, pp. 1989--2001, 2013.

\bibitem{chen2022isac}
Y.~Chen, H.~Hua, J.~Xu, and D.~W.~K. Ng, ``{ISAC} meets {SWIPT}:
  {Multi-functional} wireless systems integrating sensing, communication, and
  powering,'' \emph{arXiv preprint arXiv:2211.10605}, 2022.

\bibitem{zhang2023multi}
Y.~Zhang, S.~Aditya, and B.~Clerckx, ``Multi-functional {OFDM} signal design
  for integrated sensing, communications, and power transfer,'' \emph{arXiv
  preprint arXiv:2311.00104}, 2023.

\bibitem{li2023multi}
X.~Li, X.~Yi, Z.~Zhou, K.~Han, Z.~Han, and Y.~Gong, ``Multi-user beamforming
  design for integrating sensing, communications, and power transfer,'' in
  \emph{Proc. IEEE WCNC}, Glasgow, UK, 2023.

\bibitem{zeng2022beamforming}
X.~Zeng, L.~Xing, Y.~Wu, and Y.~Shi, ``Beamforming design for integrated
  sensing and {SWIPT} system,'' in \emph{Proc. IEEE PIMRC}, Kyoto, Japan, 2022.

\bibitem{stoica2007probing}
P.~Stoica, J.~Li, and Y.~Xie, ``On probing signal design for {MIMO} radar,''
  \emph{IEEE Trans. Signal Process.}, vol.~55, no.~8, pp. 4151--4161, 2007.

\bibitem{li2022wirelessly}
X.~Li, Z.~Han, Z.~Zhou, Q.~Zhang, K.~Huang, and Y.~Gong, ``Wirelessly powered
  integrated sensing and communication,'' in \emph{Proc. ACM MobiCom Workshop
  on ISCS}, Sydney, Australia, 2022.

\end{thebibliography}

\begin{IEEEbiographynophoto}{Xiaoyang Li} 
[M] (lixiaoyang@sribd.cn) is a research scientist at the Shenzhen Research Institute of Big Data, associated with the Chinese University of Hong Kong, Shenzhen. He received the B.Eng. degree from the Southern University of Science and Technology (SUSTech) in 2016 and the Ph.D. degree from The University of Hong Kong in 2020. From 2020 to 2022, he was a Presidential Distinguished Research Fellow at SUSTech. His research interests include integrated sensing-communication-computation, edge learning, and over-the-air computation. He is a recipient of the 2022 “Forbes 30 under 30”, the excellent award of the great bay area high value patent competition, and the exemplary reviewer of IEEE WCL and JII. He serves as workshop/session chairs of several IEEE major conference including IEEE ICC, ICASSP, and WCNC.
\end{IEEEbiographynophoto}

\begin{IEEEbiographynophoto}{Zidong Han} 
[M] (hanzd@sustech.edu.cn) received the B. S. degree in Information Engineering from the Shanghai Jiao Tong University n 2012 and the M. S. degree in Electrical Engineering from the National University of Singapore in 2013. He is currently pursuing the Ph.D. degree in the Harbin Institute of Technology, Shenzhen. He is a Senior Engineer (Information Technology) certified by the Human Resources and Social Security Department of Guangdong Province. His research interests include wireless communication, wireless control, wireless powered sensing, and edge intelligence.
\end{IEEEbiographynophoto}

\begin{IEEEbiographynophoto}{Guangxu Zhu} 
[M] (gxzhu@sribd.cn) received the Ph.D. degree in electrical and electronic engineering from The University of Hong Kong in 2019. Currently he is a research scientist at the Shenzhen research institute of big data, associated with the Chinese University of Hong Kong, Shenzhen. His recent research interests include edge intelligence, machine learning in wireless communication, federated learning, integrated sensing and communication. He is a recipient of the 2023 IEEE ComSoc Asia-Pacific Best Young Researcher Award and Outstanding Paper Award, the World's Top 2\% Scientists by Stanford University, the 2022 “AI 2000 Most Influential Scholar Award Honorable Mention”, the Young Scientist Award from UCOM 2023. He serves as associated editors at top-tier journals in ComSoc, including IEEE TWC and IEEE WCL.
\end{IEEEbiographynophoto}

\begin{IEEEbiographynophoto}{Yuanming Shi}
[SM] (shiym@shanghaitech.edu.cn) received the B.S. degree in electronic engineering from Tsinghua University, Beijing, China, in 2011. He received the Ph.D. degree in electronic and computer engineering from The Hong Kong University of Science and Technology, in 2015. Since September 2015, he has been with the School of Information Science and Technology in ShanghaiTech University, where he is currently a tenured Associate Professor. His research areas include edge AI, wireless communications, and satellite networks. He was a recipient of the IEEE Marconi Prize Paper Award in Wireless Communications in 2016, the Young Author Best Paper Award by the IEEE Signal Processing Society in 2016, the IEEE ComSoc Asia-Pacific Outstanding Young Researcher Award in 2021, and the Chinese Institute of Electronics First Prize in Natural Science in 2022. He is an IET Fellow. 
\end{IEEEbiographynophoto}

\begin{IEEEbiographynophoto}{Jie Xu} 
[SM] (xujie@cuhk.edu.cn) is currently an Associate Professor (Tenured) with the School of Science and Engineering and the Future Network of Intelligence Institute, The Chinese University of Hong Kong, Shenzhen. His research interests include wireless communications, wireless power transfer, UAV communications, edge computing and intelligence, and integrated sensing and communication. He was a recipient of the 2017 IEEE Signal Processing Society Young Author Best Paper Award, the 2019 IEEE Communications Society Asia-Pacific Outstanding Young Researcher Award, and the 2019 Wireless Communications Technical Committee Outstanding Young Researcher Award. He is the Vice Chair of the IEEE Wireless Communications Technical Committee, and the Vice Co-chair of the IEEE Emerging Technology Initiative on ISAC. He served or is serving as an Associate Editor-in-Chief of the IEEE Transactions on Mobile Computing, an Editor of IEEE Transactions on Wireless Communications, IEEE Transactions on Communications, and IEEE Wireless Communications Letters.
\end{IEEEbiographynophoto}

\begin{IEEEbiographynophoto}{Yi Gong} 
[SM] (gongy@sustech.edu.cn) received the B.Eng. and M.Eng. degrees from the Southeast University and the Ph.D. degree from the Hong Kong University of Science and Technology. He was with the Hong Kong Applied Science and Technology Research Institute, Hong Kong, and Nanyang Technological University, Singapore. He is currently a Professor with the Southern University of Science and Technology, Shenzhen, China. His research interests include cellular networks, mobile computing, and signal processing for wireless communications and related applications. He was on the Editorial Board of the IEEE Transactions on Wireless Communications and the IEEE Transactions on Vehicular Technology.
\end{IEEEbiographynophoto}

\begin{IEEEbiographynophoto}{Qinyu Zhang} 
[SM] (zqy@hit.edu.cn) received the bachelor’s degree from the Harbin Institute of Technology (HIT) in 1994, and the Ph.D. degree from the University of Tokushima, Japan, in 2003. Since 2005, he has been a Full Professor and works as the Dean of the Electronic and Information Engineering School. He is the founder and director of the Guangdong Provincial Key Laboratory of Aerospace Communication and Networking Technology. His research interests include aerospace communications and networks, wireless communications and networks, cognitive radios, signal processing, and biomedical engineering. He has received the National Science Fund for Distinguished Young Scholars, the Young and Middle-Aged Leading Scientist of China, and the Chinese New Century Excellent Talents in University. He was the Founding Chair of the IEEE Communications Society Shenzhen Chapter. He is on the Editorial Board of the Journal on Communications, KSII Transactions on Internet and Information Systems, and Science China: Information Sciences.
\end{IEEEbiographynophoto}

\begin{IEEEbiographynophoto}{Kaibin Huang} 
[F] (huangkb@eee.hku.hk) is a Professor at the Department of Electrical and Electronic Engineering, The University of Hong Kong, Hong Kong. His research interests include mobile edge computing, edge AI, and 6G systems. He received the IEEE Communication Society’s 2021 Best Survey Paper, 2019 Best Tutorial Paper, and two Asia Pacific Outstanding Paper Awards in 2015 and 2019. He is an Executive Editor of IEEE Transactions on Wireless Communications, and an Area Editor for IEEE Transactions on Machine Learning in Communications and Networking and IEEE Transactions on Green Communications and Networking. He is an IEEE Distinguished Lecturer.
\end{IEEEbiographynophoto}

\begin{IEEEbiographynophoto}{Khaled B. Letaief} 
[F] (eekhaled@ust.hk) received his Ph.D. degree from Purdue University. He has been with The Hong Kong University of Science and Technology since 1993 where he was the Dean of Engineering, and is now a Chair Professor and the New Bright Professor of Engineering. He is an ISI Highly Cited Researcher and a recipient of many distinguished awards, including the 2021 IEEE Communications Society Best Survey Paper Award, the 2019 Distinguished Research Excellence Award by the HKUST School of Engineering. He is a member of the United States National Academy of Engineering and the Hong Kong Academy of Engineering Sciences, and a fellow of the Hong Kong Institution of Engineers. He has served in many IEEE leadership positions including ComSoc President, Vice-President for Technical Activities, and Vice-President for Conferences. He is the Founding Editor-in-Chief of the prestigious IEEE Transactions on Wireless Communications and been involved in organizing many flagship international conferences.
\end{IEEEbiographynophoto}

\end{document}